\documentclass[prb,twocolumn,superscriptaddress, showpacs,floatfix]{revtex4-1}  
\usepackage{hyperref}
\usepackage{amssymb}
\usepackage{amsmath}
\usepackage{float}
\usepackage[usenames]{color}
\usepackage{graphicx}
\usepackage{epstopdf}

\begin{document}
\title{Local Hamiltonians for quantitative Green's function embedding methods}
\author{Alexander A. Rusakov}
\email[Corresponding author: ]{rusakov@umich.edu}
\affiliation{Department of Chemistry, University of Michigan, Ann Arbor, Michigan 48109, USA}
\author{Jordan J. Phillips}
\affiliation{Department of Chemistry, University of Michigan, Ann Arbor, Michigan 48109, USA}
\author{Dominika Zgid}
\affiliation{Department of Chemistry, University of Michigan, Ann Arbor, Michigan 48109, USA}

\begin{abstract}
Embedding calculations that find approximate solutions to the Schr\"{o}dinger equation for large molecules and realistic solids are performed commonly in a three step procedure involving (i) construction of a model system with effective interactions approximating the low energy physics of the initial realistic system, (ii) mapping the model system onto an impurity Hamiltonian, and (iii) solving the impurity problem.
We have developed a novel procedure for parametrizing the impurity Hamiltonian that avoids  the mathematically uncontrolled step of constructing the low energy model system.
Instead, the impurity Hamiltonian is immediately parametrized to recover the self-energy of the realistic system in the limit of high frequencies or short time. The effective interactions parametrizing the fictitious impurity Hamiltonian are local to the embedded regions, and include all the non-local interactions present in the original realistic Hamiltonian in an implicit way. We show that this impurity Hamiltonian can lead to excellent total energies and self-energies that approximate the quantities of the initial realistic system very well. Moreover, we show that as long as the effective impurity Hamiltonian parametrization is designed to recover the self-energy of the initial realistic system for high frequencies, we can expect a good total energy and self-energy. Finally, we propose two practical ways of evaluating effective integrals for parametrizing impurity models.
\end{abstract}
\maketitle
\section{Introduction}

Reliable, controlled and systematically improvable calculations for extended systems still remain a formidable task for current {\em ab initio} quantum chemistry methods. While significant progress has been made in modeling weakly correlated extended systems mostly due to various implementations of M\o ller-Plesset perturbation theory (MP2)~\cite{PhysRev.46.618}, 
 the random phase approximation (RPA)~\cite{RPA1,RPA2,RPA3,RPA_GellMann_Brueckner_1957} and coupled cluster (CC)~\cite{:/content/aip/journal/jcp/45/11/10.1063/1.1727484}, at present there is no {\em ab initio} theory that can reliably and accurately treat strongly correlated solids with $d$- and $f$-electrons in an all orbital formulation. A viable route  for these systems, that remains computationally affordable, is via embedding methods such as dynamical mean field theory (DMFT)~\cite{Georges96,Kotliar06,Metzner89,Georges92b,Held07,antoine_notes,Physics_Today}, density matrix embedding (DMET)~\cite{PhysRevLett.109.186404,doi:10.1021/ct301044e,PhysRevB.89.035140} or wave function in density functional theory (DFT) embedding~\cite{huang:194104,huang:154110,huang:084102,kluner:42,govind:7677,goodpaster:164108,goodpaster:084103,:/content/aip/journal/jcp/140/18/10.1063/1.4864040}.

In these methods the entire computationally intractable system is mapped onto an auxiliary  impurity model of strongly correlated orbitals embedded in a bath of non-interacting electrons.  The solution of the computationally tractable impurity model provides information about the local quantities of interest, such as the local Green's function or local density.
Consequently, the mapping from the infinite  system to the  impurity model is a crucial part of an embedding procedure,  one that  controls the  accuracy of the results. 
Compared to the entire  system, the impurity is described by only a few one-body and two-body body parameters.  
All non-local Coulomb interactions (represented by parameters with at least one index pointing to an orbital outside the impurity) are neglected during the construction of the impurity model. The remaining parameters have to be chosen such that the values of local impurity quantities match the local quantities of the entire system. Thus, while it is easy to define that in a Green's function embedding method an ideal set of impurity parameters should recover the local self-energy of the system, it is a much more difficult question how to find such a set of parameters. 

Multiple prescriptions have been proposed in condensed matter physics and materials science for the calculation of effective embedding interactions, $U$.
Constrained LDA (cLDA), now a standard tool for the evaluation of effective Coulomb interactions, was introduced by Dedeichs {\em et al.}~\cite{PhysRevLett.53.2512} and 
subsequently by Hybertsen {\em et al.}~\cite{PhysRevB.39.9028}
Later,  a self-consistent method for the calculation of effective interactions based on linear response within the cLDA scheme was designed 
by Cococcioni and de Gironocoli~\cite{dftu_lr}. 
This method resulted in many applications, since the calculated effective interactions were used in the computationally affordable LDA+U~\cite{dftu_1,dftu_2} method.  Aryasetiawan {\em et al.}~\cite{PhysRevB.70.195104,PhysRevB.57.4364} used the constrained random-phase approximation (RPA)~\cite{PhysRevB.70.195104,PhysRevB.77.085122,PhysRevB.80.155134} 
to exclude any screening channels and to take into account dynamical or frequency-dependent screening effects. One of the most recent advances in the field was introduced by Sch\"uler {\em et al.} ~\cite{PhysRevLett.111.036601} who proposed deriving effective interactions from the Peierls-Feynman-Bogoliubov variational principle~\cite{Peierls,Bogolyubov,Feynman}.

Conceptually, all these methods map a realistic system described by a Hamiltonian with non-local interactions onto a simpler effective model Hamiltonian with only local interactions that describe essential low energy physics of the realistic system. Subsequently, an embedding method can be employed to solve this  model  Hamiltonian. While conceptually appealing, this procedure is inherently burdened with an uncontrolled error  acquired during the mapping to the effective model Hamiltonian, and as a result,  the local impurity self-energy obtained from the embedding does not necessarily recover the local self-energy of the full realistic system.

In this paper, we introduce a different method for parametrizing the impurity model that avoids the issue of mapping to the effective Hamiltonian. We postulate that a method for finding effective Coulomb interactions should be designed to approximate either the local GreenÕs function or equivalently self-energy of the full realistic system, thus providing a well defined mathematical criterion for finding the effective interactions. We propose an approach for finding the effective Coulomb interactions that is designed such that the impurity model recovers the frequency dependent self-energy of the full system in the high frequency limit. Our prescription for finding the effective Coulomb interactions is mathematically well defined and completely general.

While the most obvious use of our procedure is for embedding methods such as DMFT, we do not attempt such a study in this paper 
because the embedding method itself can introduce an error. Here we only aim to calibrate the approximation resulting from the use of the effective interactions. To this end we have designed several tests that measure  the accuracy of our impurity parametrization. First, we compare the electronic energy from our procedure to that of prototypical systems for which we are able to obtain an exact energy and self-energy. Second, since for multi-orbital impurities there is no single unique parametrization of effective interactions, we will investigate if different parametrizations recover similar energetics. Lastly, we will establish if our parametrization, which recovers the self-energy of the full system in the high frequency limit, yields an acceptable impurity self-energy in the low frequency limit when compared with the exact answer.

The current paper is organized as follows: In Sec.~\ref{motivation} and \ref{procedure} we discuss the scheme for evaluating  effective Coulomb interactions. In Sec.~\ref{results} we show the calibration results and  compare them to the exact results.  In Sec.~\ref{general} we discuss the generalization of the procedure to larger systems and present necessary calibrations. Finally, Sec.~\ref{conclusions} presents the overall conclusions of our work.

\section{Effective interactions based on the high frequency expansion of the self-energy}\label{motivation}
We define a general Hamiltonian 
\begin{equation}\label{ham}
\hat{H}=\sum^{n}_{ij}t_{ij}a^{\dagger}_{i}a_{j}+\frac{1}{2}\sum^{n}_{ijkl}v_{ijkl}a^{\dagger}_{i}a^{\dagger}_{k}a_{l}a_{j},
\end{equation}
for a realistic system (a molecule or a solid) with full non-local Coulomb interactions (in chemists' notation) between all $n$ orbitals 
\begin{equation}
v_{ijkl}=\iint  d{\bf r}_1 d{\bf r}_2 \phi^{*}_i({\bf r}_1 ) \phi_j({\bf r}_1 )\frac{1}{r_{12}}\phi^{*}_k({\bf r}_2 ) \phi_l({\bf r}_2 )
\end{equation}
and one-body operators
\begin{equation}
t_{ij}=\int d{\bf r}_1 \phi^{*}_i({\bf r}_1) h({\bf r}_1)\phi_j({\bf r}_1 ),
\end{equation}
\begin{equation}
h({\bf r}_1)=-\frac{1}{2}\nabla^2 _{{\bf r}_1} -\sum_A \frac{Z_A}{|{\bf r}_1-{\bf R_A}|}.
\end{equation}
The correlated Green's function $G(\omega)$ for this system is related to the non-interacting Green's function $G^0(\omega)$ via the Dyson equation
\begin{equation}
\Sigma_{\infty}+\Sigma(\omega)=[G^0(\omega)]^{-1}-[G(\omega)]^{-1},
\end{equation}
where $\Sigma_{\infty}$ and $\Sigma(\omega)$ are the frequency independent and frequency dependent parts of the self-energy, which describe all correlation effects present in the realistic Hamiltonian in Eq.~\ref{ham}.

Imagine now that in our molecule or solid we choose a subset of orbitals, called  the correlated local subspace, which we deem important for the physical description of this system. Then we can express both parts of the self-energy as a sum of local and non-local contributions
\begin{eqnarray}
&\Sigma_{\infty}&=\Sigma^{loc}_{\infty}+\Sigma^{non-loc}_{\infty},\\
&\Sigma(\omega)&=\Sigma^{loc}(\omega)+\Sigma^{non-loc}(\omega),
\end{eqnarray}
where the local contributions come from the embedding calculations for the correlated local subspace.

The calculation of $\Sigma_{\infty}$, corresponding to the (frequency independent) Hartree-Fock (HF) self-energy,  is usually computationally affordable since it scales polynomially and requires only $O(n^4)$ operations. In practical embedding calculations, $\Sigma^{loc}_{\infty}$ is constructed using the correlated subspace integrals multiplied with the correlated density matrix, while $\Sigma^{non-loc}_{\infty}$ is usually approximated at the HF or DFT level by multiplying the HF/DFT density matrix with all the remaining integrals~\cite{Zgid11,PhysRevLett.106.096402,Kotliar06,Held07}.

The frequency dependent self-energy, $\Sigma(\omega)$, contains the important many-body effects. 
In embedding calculations the $\Sigma^{loc}(\omega)$ part of  this self-energy is evaluated by solving a simpler Hamiltonian representing a fictitious system, where the Hamiltonian 
 is constructed to recover the local Green's function and self-energy of the realistic system. This Hamiltonian has  effective  two-body interactions given by $U_{ijkl}\ne 0$ if all orbital indices belong to the correlated subspace, and $U_{ijkl}= 0$ if at least one of the indices is outside the correlated subspace.
The non-local frequency dependent part of the self-energy, $\Sigma^{non-loc}(\omega)$,  cannot be recovered for orbitals outside the correlated subspace by frequency independent methods such as HF or DFT. Rather, in these methods $\Sigma^{non-loc}(\omega)$ is simply zero.
Consequently, the total self-energy can be written as
\begin{equation}
\Sigma^{embed}=\Sigma^{loc \ embed}_{\infty}+\Sigma^{non-loc\ embed}_{\infty}+\Sigma^{loc \ embed}(\omega).
\end{equation}

We would like the embedding calculation to approximate in the best possible way the local quantities for the full system. Consequently, the self-energy for the full system has to be approximated by the following self-energies:
\begin{eqnarray}
&\Sigma^{full} &\approx \Sigma^{embed},\\
&\Sigma^{full}_{\infty}& \approx \Sigma^{loc \ embed}_{\infty}+\Sigma^{non-loc\ embed}_{\infty},\\\label{sigma_infty_recovery}
&\Sigma^{full}(\omega)& \approx \Sigma^{loc \ embed}(\omega),\label{sigma_freq_recovery}
\end{eqnarray}
where in the last equation we used $\Sigma^{non-loc}(\omega)=0$. Since $\Sigma^{embed}_{\infty}$ has both local and non-local parts, let us assume that it is a good approximation to the $\Sigma^{full}_{\infty}$ of the full system. 
Because the frequency dependent self-energy of the full system from Eq.~\ref{sigma_freq_recovery} should be recovered only by the local self-energy coming from the embedded orbitals, we should find mathematical conditions which will ensure that $ \Sigma^{loc \ embed}(\omega)$ reasonably approximates the frequency dependent self-energy of the full system.

In general frequency dependent effective interactions, $U(\omega)$, are required to find a solution that fulfills Eq.~\ref{sigma_freq_recovery}
for every frequency. However frequency independent  effective interactions, $U$, are sufficient to enforce the equality of the self-energies in the high frequency limit. The recovery of this limit is important to describe the short time behavior of the Green's function, and an accurate computational method should recover at least this limit of the self-energy. To find a set of effective interactions that fulfill Eq.~\ref{sigma_freq_recovery} in the high frequency limit, we start with analyzing
the high frequency expansion of the Green's function~\cite{Comanac}
\begin{equation}
G(i\omega)=\frac{G_1}{i\omega}+\frac{G_2}{(i\omega)^2}+\frac{G_3}{(i\omega)^3}+O\left(\frac{1}{(i\omega)^4}\right), 
\end{equation}
or in general 
\begin{equation}
[G(i\omega)]_{ij}=\sum_{k\ge 0} (-1)^{(k-1)} \frac{\langle \Psi_m| \{[\hat{H},a_{i}]_{\{k\}},a^{\dagger}_{j}\}|\Psi_m\rangle}{(i\omega)^k}.
\end{equation}
In the numerator of the above equation the commutator is defined as  $[\hat{H},a_{i}]_{\{k\}} = \underbrace{[\hat{H},[\hat{H},[...[\hat{H},a_{i}]]...]]}_{k\:operators\:totally}$ with $|\Psi_m\rangle$ being the 
solution of the Schr\"odingier equation  $\hat{H}|\Psi_m\rangle=E_m|\Psi_m\rangle$ for the Hamiltonian in Eq.~\ref{ham}. 

Analogously, we can write the high frequency expansion of the self-energy as
\begin{equation}
\Sigma(i\omega)=\Sigma_{\infty}+\frac{\Sigma_1}{i\omega}+\frac{\Sigma_2}{(i\omega)^2}+\frac{\Sigma_3}{(i\omega)^3}+O\left(\frac{1}{(i\omega)^4}\right).
\end{equation}
Using the Dyson equation we can then evaluate the coefficients of the self-energy expansion as
\begin{eqnarray}
&\Sigma_{\infty}&=G_2 - G^{0}_2, \\
&\Sigma_1&=(G^{0}_2)^2- (G_2)^2+ G_3-G^{0}_3 \label{sigma1}.
\end{eqnarray}
Enforcing Eq.~\ref{sigma_freq_recovery} requires matching of the full and embedded system's self-energy at least  up to the first order in $1/\omega$ in the high frequency limit
\begin{equation}\label{sigma1_match}
\Sigma^{full}_1=\Sigma^{loc \ embed}_1.
\end{equation}
A general expression for $\Sigma_1$ is given by Eq.~\ref{sigma1}  with the second and third coefficient in the Green's function expansion 
\begin{eqnarray}
&[G_2]_{ij}&=t_{ij}+\sum_{rs}\gamma_{rs}(v_{ijrs}-\frac{1}{2}v_{isrj}), \\\label{G2}
&[G_3]_{ij}&=\sum_{l}t_{il}t_{lj}+\sum_{qrs}v_{iqrs}(t_{qj}\gamma_{rs}-\frac{1}{2}t_{sj}\gamma_{rq})\\\nonumber \label{G3}
&+&\sum_{qrs}t_{ir}\gamma_{qs}(v_{qsrj}-\frac{1}{2}v_{qjrs})\\ \nonumber
&-&\sum_{klqrs}v_{qrkl}\gamma_{qksl}(v_{ijrs}-\frac{1}{2}v_{isrj})\\ \nonumber
&+&\frac{1}{2}\sum_{klqrs}v_{qrkj}v_{ilrs}\gamma_{qksl}\\ \nonumber
&+&\sum_{lqrs}v_{iqrs}\gamma_{rl}(v_{qjsl}-\frac{1}{2}v_{qlsj})\\\nonumber
&+&\sum_{klqrs}v_{iqrs}\gamma_{rksl}(v_{qjkl}-\frac{1}{2}v_{qlkj})\\\nonumber
&-&\frac{1}{2}\sum_{klqrs}v_{qskl}v_{iqrj}\gamma_{rksl}\\\nonumber
&-&\frac{1}{2}\sum_{klqrs}v_{iqrs}(v_{sjkl}\gamma_{rkql}+v_{slkj}\gamma_{rklq})\\\nonumber
&+&\sum_{klqrs}v_{sqkl}v_{ijrs}\gamma_{rkql}.\nonumber
\end{eqnarray}
In these expressions the one- and two-body density matrices are defined as
\begin{equation}
\gamma_{ij}=\langle \Psi_m |\sum_{\sigma}a^{\dagger}_{i\sigma} a_{j\sigma}|\Psi_m\rangle
\end{equation}
and 
\begin{equation}
\gamma_{ijkl}=\langle \Psi_m |\sum_{\sigma \tau}a^{\dagger}_{i\sigma}a^{\dagger}_{j\tau} a_{l\tau}a_{k\sigma}|\Psi_m\rangle,
\end{equation}
respectively.
$\Sigma^{full}_1$ for the full system is computed with all the local and non-local Coulomb interactions, and with both one- and two-body density matrices. In contrast the embedded system's $\Sigma^{loc \ embed }_1$ coefficient is computed by solving for the Green's function of the fictitious system (impurity) that has only local Coulomb interactions, $U_{ijkl}$ with all the indices belonging to the correlated subspace. 

While all the previous arguments were general,  for simplicity we will consider explicitly a case for a single embedded orbital with a single on-site interaction denoted by $U$. For a single embedded orbital,  the solution of the fictitious impurity Hamiltonian yields $\Sigma^{loc \ embed}_1$ expressed as
\begin{equation}\label{single_site_dmft_se}
\Sigma^{loc \ embed}_1 = \frac{1}{2}U^2\gamma_{11}(1-\frac{1}{2}\gamma_{11}).
\end{equation}
Additionally, if we assume that our calculation for the embedded system yields accurate density matrices, it is obvious that Eq.~\ref{single_site_dmft_se} cannot provide a good approximation to $\Sigma^{full}_1$  in Eq.~\ref{sigma1} and Eqs.~\ref{G2}-\ref{G3}, as these involve all the local and non-local Coulomb interactions. 
We emphasize that this discrepancy, $\Sigma^{full}_1 \neq \Sigma^{loc \ embed}_1$, is only present if the full system's Hamiltonian includes the non-local Coulomb interactions outside the correlated subspace of the embedded system. For systems such as the Hubbard lattice (with only on-site $U$ which are fully within the correlated subspace) such a difference in $\Sigma_1$ will not be observed. 

These observations lead us to an important question: how can we improve the high frequency self-energy behavior of the fictitious system with only local interactions, to make it approximate a realistic system better and account for the neglected non-local interactions? 
If we assume for simplicity  that the exact $\Sigma^{full}_1$ is known and the 
embedded orbital has only the on-site interaction, then Eq.~\ref{sigma1_match} can be trivially fulfilled by adjusting  the on-site $U$ while performing the following reparametrization
\begin{equation}\label{single_site_rep}
U_{eff}=\sqrt{\frac{2\Sigma_1}{\gamma_{11}(1-\frac{1}{2}\gamma_{11})}}.
\end{equation}
This reparametrization 
 will improve the high-frequency behavior of the fictitious system and will recover the local self-energy of the full system in the high frequency limit, thus providing prerequisites for a good approximation.
While $U_{eff}$ in Eq.~\ref{single_site_rep} accounts only for the on-site interactions, an extension of this procedure can be formulated to calculate a subset of effective interactions in a correlated multi-orbital subspace.
 Such a procedure will be discussed in Sec.~\ref{results}.

\section{Numerical procedure}\label{procedure}
As stated previously, since the embedding method itself can introduce an error, here we only  calibrate the approximation resulting from the use of the effective interactions in this paper.
Consequently we embed a subset of correlated orbitals with effective local interactions, $U$, into a set of orbitals described by full configuration interaction (FCI). 
This is achieved  by employing the following definition for the zero-order Green's function
\begin{equation}\label{G01}
G^0(\omega)=[(\omega+\mu)-\bar{F}]^{-1},
\end{equation}
where $\mu$ is the chemical potential and 
\begin{equation}\label{fock_m}
\bar{F}_{ij}=F_{ij}-\Sigma^{loc \ embed}_\infty,
\end{equation}
with the local part of self-energy for embedded orbitals defined as
\begin{equation}\label{sigmalocjord}
\Sigma^{loc \ embed}_\infty=\sum_{kl \in loc}\gamma_{kl}(U_{ijkl}-\frac{1}{2}U_{ilkj}),
\end{equation}
where the sum runs over orbitals from the local correlated embedded subspace.
The Fock matrix, $F_{ij}$, is defined as $F_{ij}= t_{ij}+[\Sigma_\infty]_{ij}$, with $\Sigma_\infty$ evaluated using the correlated density matrix  from FCI calculations. The correlated one-body density matrix $\gamma$ from Eq.~\ref{sigmalocjord} comes from calculations with the correlated orbitals parametrized with effective interactions, $U_{ijkl}$. 
This prescription ensures that the only error in our calculations can result from a wrong self-energy in the low frequency limit caused by the parametrization of effective integrals based on the high frequency expansion.

The definition of the zero-order Green's function from Eq.~\ref{G01}~and~\ref{fock_m} assumes that the correlated Green's function of the full system is represented as
\begin{equation}\label{corr_gf}
G(\omega)=[(\omega+\mu)-\bar{F}-\Sigma^{loc \ embed}_\infty-\Sigma^{loc \ embed}(\omega)]^{-1}.
\end{equation}

Before we discuss the numerical results, let us define the details of the scheme we are employing to calibrate the accuracy. 
We apply our procedure for finding effective interactions to the H$_{6}$ ring with a regular hexagonal arrangement of atoms. Our calculations are performed in small STO-6G and double-zeta (DZ) basis sets since  FCI results can be readily obtained and exact $\Sigma_{\infty}$, $ \Sigma(\omega)$, and $\Sigma_{1}$ matrices can be explicitly computed. These exact quantities will be used for the comparison against our results.
To disentangle the embedding error from the error of parametrization of effective integrals, we define the fictitious system used to evaluate the frequency dependent self-energy as a ring with only the on-site interactions $U_{iiii}$, where all the remaining interactions will be used to construct $\Sigma^{non-loc \ embed}_\infty$ with a FCI density.
Since the density matrix for the embedded orbital is being adjusted during our calculations, we employ the following self-consistency scheme:
\begin{center}
\bf{Iterative Scheme I}
\end{center}
\begin{enumerate}
\item perform a FCI calculation on the entire system with all the Coulomb interactions $v_{ijkl}$
\item choose local orbitals that should be embedded
\item compute $\Sigma^{full}_{1}$ from FCI for the entire system
\item compute one-body density matrix, $\gamma^{(1)}$, and two-body density matrix, $\gamma^{(2)}$, from FCI
\item  compute local $U_{ijkl}$ for the embedded orbitals using $\Sigma^{full}_{1}$ and the density matrices $\gamma^{(1)}$ and $\gamma^{(2)}$
\item calculate $\bar{F}$ from Eq.~\ref{fock_m} using $\gamma^{(1)}$
\item compute the self-energy for the embedded orbitals as $\Sigma^{loc \ embed}_{\infty}+\Sigma^{loc \ embed}(\omega)$
\item compute new correlated Green's function for the entire system from Eq.~\ref{corr_gf}
\item update $\gamma^{(1)}$~\footnote{in general, we should also update $\gamma^{(2)}$ here, although for the sake of simplicity we refrain from doing so}
\item evaluate electronic energy for the entire system
\item go to step 5 until convergence
\end{enumerate}
In all our calculations we use a grid of 3000 Matsubara frequencies on the imaginary axis, with an inverse temperature $\beta = 50$.

\section{Numerical results}\label{results}
\subsection{Accuracy of electronic energies}

First let us define effective integrals for our calibration studies. We embed either (i) a single orbital or (ii) two orbitals. 
If a single orbital is embedded then our fictitious system is parametrized by the on-site effective integral, $U_{iiii}$, where $i$ is the orbital index. This effective on-site interaction is uniquely defined by Eq.~\ref{single_site_rep} since there is only one element, $[\Sigma^{full}_1]_{ii}$, and a single on-site  $U_{iiii}$.
For two embedded orbitals the fictitious system has effective interactions local to the two subspace orbitals. In this subspace there are four distinct (up to permutations) bare integrals $v_{1111}$, $v_{1122}$, $v_{1212}$, and $v_{1112}$ (due to symmetry, $v_{1111} = v_{2222}$ and $v_{1112} = v_{2221}$) and only two unique elements of the $\Sigma^{full}_1$ matrix.  Consequently, multiple choices of effective integrals of $U_{ijkl}$ are allowed to parametrize elements $[\Sigma^{full}_1]_{11}$ and $[\Sigma^{full}_1]_{12}$ simultaneously. 
The simplest approach is to re-scale uniformly  all bare integrals, $U_{ijkl} = \alpha \cdot v_{ijkl}$, by a scaling parameter $\alpha$ chosen to fit $\Sigma^{full}_{1}$ best  in the least-square sense. An approximation frequently resulting in a better  $\Sigma^{full}_{1}$ can be attained by introducing more scaling parameters, though one should be aware of
potential optimization stability issues if the number of parameters exceeds the number of independent $\Sigma^{full}_{1}$ elements. In the present case, the problem remains well-posed if two parameters are introduced, for instance: $U_{iiii} = \alpha \cdot v_{iiii}, i = 1, 2$ and $U_{ijkl} = \beta \cdot v_{ijkl}$ for other $ijkl$ combinations. We have also attempted to introduce four parameters, \textit{i.e.} to scale each class of integrals independently: 
$U_{1111} = \alpha \cdot v_{1111}$, $U_{1122} = \beta \cdot v_{1122}$, $U_{1212} = \gamma \cdot v_{1212}$, and $U_{1112} = \delta \cdot v_{1112}$. Since there can be multiple sets of such $\alpha$, $\beta$,
$\gamma$, and $\delta$ parameters, we have restricted the search by imposing the Schwarz inequality to retain ``physically meaningful'' values of effective integrals. In addition, $U_{ijkl}$ resulting from the two-parameter scaling served as the initial guess for the optimization procedure. 
These parametrizations  are presented in Table~\ref{various_param} for several bond distances.
\begin{table}[h!]
\caption{\label{various_param} Possible effective integrals for H$_{6}$ in the STO-6G basis and corresponding total energies. $n$o/$m$p stands for $n$ orbitals and $m$ scaling parameters. The parametrizations denoted with a star contain effective integrals obtained from a fully self consistent iterative scheme.}
\begin{tabular}{|c|c|c|c|c|c|c|}
\hline
 type & $U_{1111}$ & $U_{1122}$ & $U_{1212}$ & $U_{1222}$ & E, a.u. \\
\hline
\multicolumn{6}{|c|}{$d = 1.4$ a.u.} \\
\hline
 1o/1p & 0.5984 &--- &--- & --- & $-3.0665$\\
 2o/1p & 0.6182 & 0.3533 & $0.0093$ & $-0.0062$ &$-3.0557$	\\
 2o/2p & 0.6378 & 0.1266 & $0.0033 $ & $-0.0022 $ &$-3.0641$	\\
 2o/2p* & 0.6274 &$0.0817 $ & $0.0022$ & $-0.0014$ &$-3.0652$	\\	
 2o/4p* & 0.6379 &0.1332 & $0.0002$ & $0.0008$ &$-3.0640$	\\
 \hline
\multicolumn{6}{|c|}{$d = 1.8$ a.u.} \\
\hline
 1o/1p & 0.5952 &--- &--- & --- & $-3.2583$\\
 2o/1p & 0.6235 & 0.3290 & $0.0077$ & $-0.0059$ &$-3.2471$	\\
 2o/2p & 0.6320 & 0.1275 & $0.0030 $ & $-0.0023 $ &$-3.2555$	\\
 2o/2p* & 0.6190 &$0.0699 $ & $0.0016$ & $-0.0013$ &$-3.2571$	\\	
 2o/4p* & 0.6303 &0.1260 & $0.0000$ & $0.0041$ &$-3.2556$	\\
\hline
\multicolumn{6}{|c|}{$d = 2.4$ a.u.} \\
\hline
 1o/1p & 0.6283 &--- &--- & --- & $-3.1589$\\
 2o/1p & 0.6614 & 0.3060 & $0.0054$ & $-0.0067$ &$-3.1529$	\\
 2o/2p & 0.6634 & 0.1688 & $0.0030 $ & $-0.0037 $ &$-3.1561$	\\
 2o/2p* & 0.6401 &$0.0400 $ & $0.0007$ & $-0.0009$ &$-3.1581$	\\	
 2o/4p* & 0.6253 &0.0128 & $-0.0099$ & $0.0124$ &$-3.1591$	\\
\hline
\multicolumn{6}{|c|}{$d = 3.4$ a.u.} \\
\hline
 1o/1p & 0.7290 &--- &--- & --- & $-2.9122$\\
 2o/1p & 0.7476 & 0.2730 & $0.0022$ & $-0.0072$ &$-2.9352$	\\
 2o/2p & 0.7440 & 0.4191 & $0.0034 $ & $-0.0111 $ &$-2.9374$	\\
 2o/2p* & 0.7419 &$0.2709 $ & $0.0022$ & $-0.0097$ &$-2.9295$	\\	
 2o/4p* & 0.7387 &0.0786 & $0.0043$ & $-0.0376$ &$-2.9230$	\\
\hline
\multicolumn{6}{|c|}{$d = 4.0$ a.u.} \\
\hline
 1o/1p & 0.7593 &--- &--- & --- & $-2.8500$\\
 2o/1p & 0.7675 & 0.2439 & $0.0011$ & $-0.0062$ &$-2.8714$	\\
 2o/2p & 0.7678 & 0.2641 & $0.0011 $ & $-0.0067 $ &$-2.8720$	\\
 2o/2p* & 0.7655 &$0.2432 $ & $0.0011$ & $-0.0067$ &$-2.8689$	\\	
 2o/4p* & 0.7618 &0.0776 & $0.0006$ & $-0.0228$ &$-2.8630$	\\ 
\hline
\end{tabular}
\end{table}

\begin{figure}[her!]
\centering
\includegraphics[width=1.0\columnwidth]{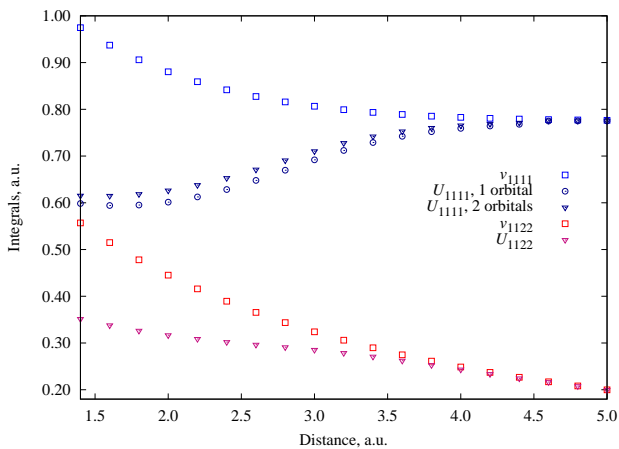}
\caption{\label{screened}  Bare $v$ and screened $U$ integrals as a function of bond distance for one and two embedded orbitals for the H$_{6}$ ring in the STO-6G basis. The distance on the x-axis is the radius of the H$_{6}$ ring molecule (distance from the center of the ring to the hydrogen atom nucleus).}
\end{figure}

Additionally  in Fig.~\ref{screened} for both cases of a single embedded orbital and two embedded orbitals we explicitly plot effective integrals, $U_{ijkl}$, as compared to bare integrals, $v_{ijkl}$. 
Note that our orbitals and integrals are in the orthogonalized basis rather than in the initial non-orthogonal Gaussian atomic orbital basis set. Because the orthogonalization is performed via
the L\"{o}wdin transformation involving an overlap matrix, the on-site integrals $v_{1111}$ are not a constant function of the interatomic distance. 
As expected, the deviations of effective integrals $U$ from bare Coulomb integrals $v$  (which is sometimes called "screening") vanishes at dissociation, but manifests itself clearly for shorter bond distances where non-local Coulomb integrals become significant.
We interpret the screening of the local two-electron integrals as a mathematical feature of a local model with only on-site effective interactions that  incorporate the non-local interactions.

\begin{figure}[her!]
\includegraphics[width=1.0\columnwidth]{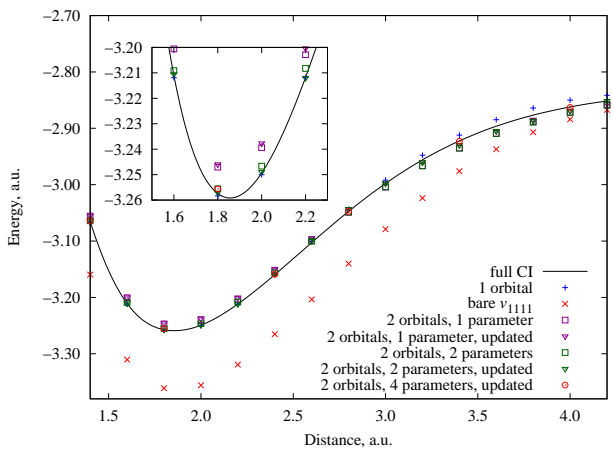}
\caption{\label{total} Total energies for various parametrizations of local two-electron integrals compared against the FCI total energy for H$_{6}$ in the STO-6G basis set as a function of bond distance.}
\end{figure}

\begin{figure}[her!]
\includegraphics[width=1.0\columnwidth]{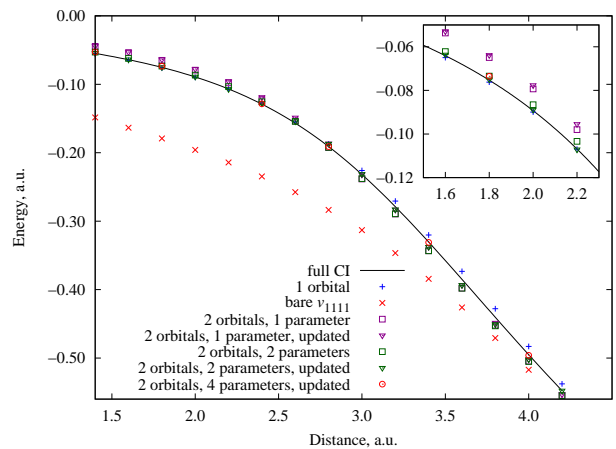}
\caption{\label{correlation} Correlation energies for various parametrizations of local two-electron integrals compared against the FCI correlation energy for H$_{6}$ in the STO-6G basis set as a function of bond distance.}
\end{figure}

After defining the effective integrals, we now investigate the accuracy of electronic energies calculated using the Galitskii-Migdal formula~\cite{GM_formula} applied to the correlated Green's function obtained from the above defined self-consistency scheme. 
For the  H$_{6}$ ring in the STO-6G basis set we present in Fig.~\ref{total} the energies evaluated using  effective integrals $U_{ijkl}$,  the energy
obtained with bare local Coulomb integrals, and the FCI energy.
The correlation energy, defined as $E_{corr} = E_{correlated} - E_{Hartree-Fock}$,  is plotted in Fig.~\ref{correlation}. 

\begin{figure}[her!]
\centering
\includegraphics[width=1.0\columnwidth]{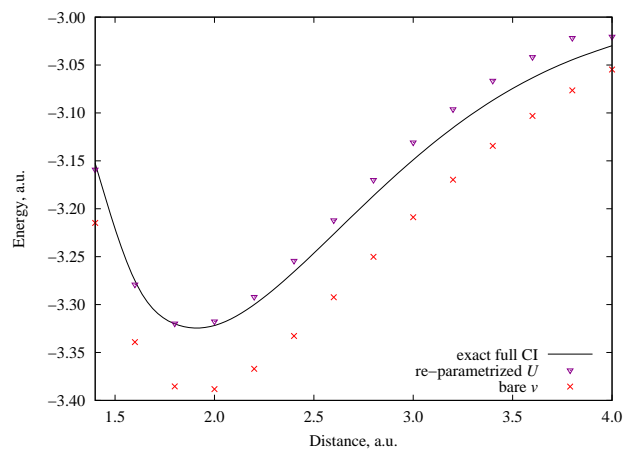}
\caption{\label{total_dz} Total energy obtained with four effective parameters in a  two-orbital subspace compared against the FCI total energy for H$_{6}$ in the DZ basis set as a function of bond distance.}
\end{figure}

\begin{figure}[her!]
\centering
\includegraphics[width=1.0\columnwidth]{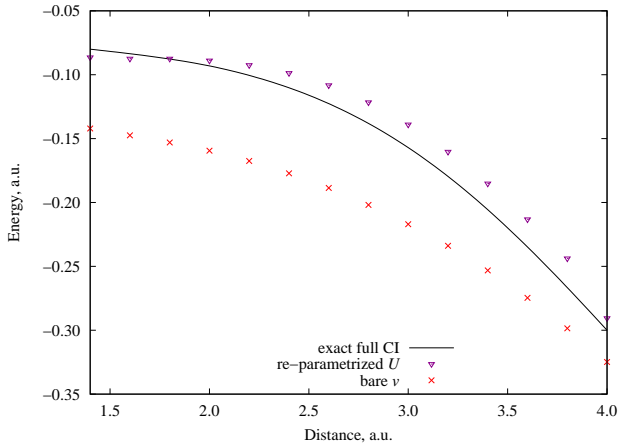}
\caption{\label{correlation_dz} Correlation energy obtained with four effective parameters in a  two-orbital subspace  compared against the FCI correlation energy for H$_{6}$ in the DZ basis set as a function of bond distance.}
\end{figure}

To investigate how our procedure is affected by enlarging the number of embedded orbitals, we also performed a study in the DZ basis. The effective integrals were used to parametrize two orbitals that are centered on every hydrogen atom.  This  involved four parameters to scale the groups of two-electron integrals and was chosen to fit, in the best way possible, the $2\times 2$ matrix of $ \Sigma^{full}_1$ for every hydrogen atom described by two orbitals. The total energies and correlation energies from this parametrization are listed in Fig.~\ref{total_dz} and Fig.~\ref{correlation_dz}, respectively.

There are several general points from analyzing Figs.~\ref{total}-\ref{correlation_dz} that should be noted. First, the fictitious Hamiltonian with on-site bare integrals yields very poor total and correlation energies anywhere away from the dissociation limit. The deviation from the exact FCI data exceeds 0.1 a.u. around equilibrium. This is not surprising though, as this fictitious Hamiltonian completely neglects non-local integrals and non-local contributions to the self-energy. Such an approximation is valid only in the dissociation limit. 

Second, all explored parametrizations of the local two-electron integrals emulating the non-local contributions to the first-order self energy $\Sigma_{1}$ lead to a drastic improvement over the case of bare integrals. Typical values of the deviations from the exact FCI data are around 0.01 a.u., which is an order of magnitude less than with bare integrals. Additionally, reproducing $\Sigma_{1}$ with high accuracy (the largest deviation from the exact $\Sigma_{1}$ elements is of  order  $10^{-4}$ a.u.) employing four parameters allows us to recover almost the exact FCI result, as shown in Table~\ref{4param} for several points along the dissociation curve. 
\begin{table}[h!]
\caption{\label{4param}H$_{6}$, Full CI energy, $E_{FCI}$ and  energy, $E_{4 \ param}$, obtained using a fictitious system Hamiltonian where four local parameters were used to describe the interactions in a two-orbital correlated subspace of the H$_6$ ring in the STO-6G basis set.}
\begin{tabular}{c|c|c}
R, a.u. & E$_{FCI}$, a.u. &  E$_{4 \ param}$, a.u.  	\\
	\hline
	\hline
1.4   &  -3.06585     &   -3.06398 \\
1.8   &  -3.25742    &    -3.25562 \\
2.4   &  -3.15968   &     -3.15910 \\
2.8  &   -3.04747     &   -3.04751 \\
3.4   &  -2.92238   &     -2.92298 \\
4.0  &   -2.86210    &    -2.86300 \\
	 \hline
\end{tabular}
\end{table}

Third, despite there being multiple ways to choose  $U_{ijkl}$ that approximate $\Sigma_{1}$, as long as such $U_{ijkl}$ reproduce the exact $\Sigma_{1}$, the resulting fictitious Hamiltonians lead to comparably good energies that approximate the exact FCI energy well. 

\subsection{Accuracy of the frequency dependent self-energy}

\begin{figure}[her!]
\centering
\includegraphics[width=1.0\columnwidth]{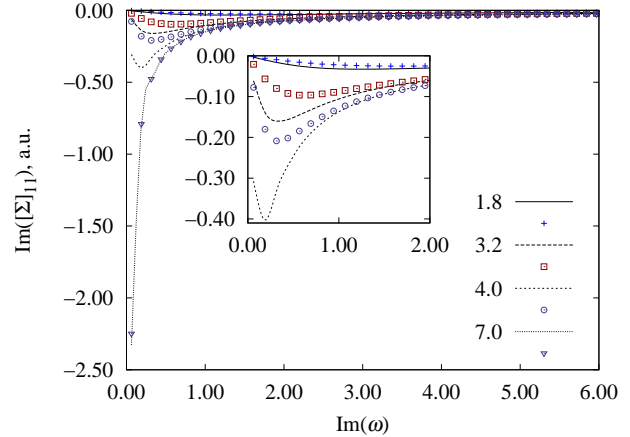}
\caption{\label{SE} The diagonal element of the exact FCI self-energy and the diagonal element of Im$(\Sigma^{loc \ embed}(\omega)$ evaluated  with four effective parameters in a  two-orbital subspace for the H$_6$ ring molecule in the STO-6G basis set. Exact values are plotted with lines, approximate values with points.} 
\end{figure}

Since our  parametrization of effective interactions was developed to approximate the high frequency self-energy expansion, we cannot expect that the full system's self-energy  $\Sigma^{full}(\omega)$ from Eq.~\ref{sigma_freq_recovery} will be recovered exactly for low frequencies. In this subsection, we calibrate the error of the low frequency self-energy. To this end,  we have examined the behavior of the self-energy $\Sigma^{loc \ embed}(\omega)$. Since the real frequency independent part $\Sigma^{non-loc \ embed}_{\infty}$ is included exactly, we consider here only 
Im$(\Sigma^{loc \ embed}(\omega))$. 
In Fig.~\ref{SE}, for several bond distances of the H$_6$ ring molecule, we compare the self-energy calculated
using four  parameters to obtain effective interactions for  a two-orbital subspace  with the FCI self-energy.
Though the self-energy for the two-orbital subspace is a $2 \times 2$ matrix for every frequency, in Fig.~\ref{SE} we plot only diagonal values, since the off-diagonal ones are at least an order of magnitude smaller 
and behave the same way.
To better visualize the low-frequency region, we plot the self-energy for the first 50 frequencies out of 3000 used in the evaluation of the Green's function. 
Though the only condition imposed on the  approximate self-energy is recovering the exact $\Sigma_{1}$ in the high-frequency limit, the resulting frequency-dependent self-energy deviates from the
exact result in a rather limited range of low frequencies.
The difference between the exact and  approximate self-energy is smallest for  short distances, where our prescription seems to be accounting for the non-local interactions not present in the fictitious model very well. 
The differences become largest for the intermediate bond distances, where our fictitious model  is not able to describe fully the low frequency behavior. 
We attribute this difference to the inability of the two-orbital type of fictitious system to emulate all the types of correlations present in the full model. 
This self-energy error contributes to a small total energy error for intermediate bond distances.

 The above observations show that the non-local contributions, captured via adjusting local integrals to fit $\Sigma^{full}_{1}$, are sufficiently dominant to yield the correct qualitative and quantitative behavior of the self-energy in the low- and high-frequency regions, respectively. To improve the self-energy in the intermediate distances, a larger number of embedded orbitals would be needed. 

\begin{figure}[her!]
\centering
\includegraphics[width=1.0\columnwidth]{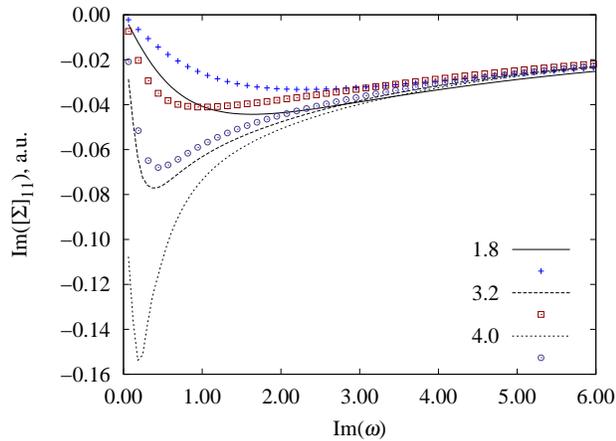}
\caption{\label{SE_dz_11} H$_{6}$, The FCI diagonal elements of the self-energy and the diagonal element of Im$(\Sigma^{loc\ embed}(\omega))$ for the H$_6$ ring in the DZ basis set. Exact values are plotted with lines, approximate values with points}
\end{figure}

\begin{figure}[her!]
\centering
\includegraphics[width=1.0\columnwidth]{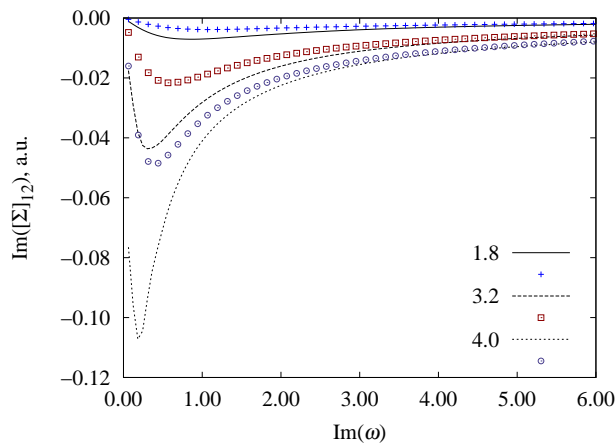}
\caption{\label{SE_dz_12}  The FCI off-diagonal elements of the self-energy and the off-diagonal element of Im$(\Sigma^{loc\ embed}(\omega))$ for the H$_6$ ring in the DZ basis set. Exact values are plotted with lines, approximate values with points}
\end{figure}

We also have investigated the self-energies for the H$_{6}$ ring in the DZ basis where the same effective integrals were used as for the energy calibration.
The imaginary parts of the diagonal and off-diagonal element of the self-energy  (Figs.~\ref{SE_dz_11},~\ref{SE_dz_12}) display the same trends as observed for the small basis, \textit{i.e.} quantitative agreement of the exact and approximate
self-energies except for a narrow range of low frequencies. 

\section{Approximating $\Sigma^{full}_1$ for large systems}\label{general}

For the purpose of our calibration study  the exact $\Sigma^{full}_1$ was known and was used to calculate effective Coulomb interactions. However in typical calculations for large realistic systems the exact $\Sigma^{full}_1$ will obviously be unknown. Therefore, it needs to be initially approximated in order to calculate effective Coulomb interactions that recover the self-energy of the full system in the high frequency limit. The simplest  approximations to the $\Sigma_1$ matrix can be obtained from
\begin{itemize}
\item  the explicit self-energy obtained from computationally affordable methods such as  the lowest order of perturbation theory expressed in Green's function language~\cite{:/content/aip/journal/jcp/140/24/10.1063/1.4884951,Dahlenjcp2005}. 
\item indirectly from methods that do not have an explicit frequency dependence such as DFT. The one- and two-body density matrices produced in these methods can be used used to calculate $\Sigma_1$ from Eqs.~\ref{sigma1}--\ref{G3}.
\end{itemize}
Here, we  discuss both options and suggest how they can be employed to calculate effective integrals.

\subsection{Approximating $\Sigma^{full}_1$ using the cumulant expansion}

Computationally affordable methods that do not exploit an explicit frequency dependence such as DFT  or MP2 can be employed to approximate $\Sigma_1$ for hundreds of orbitals. 
The one-body density matrix that is obtained in DFT or MP2 can be later used in the cumulant expansion~\cite{:/content/aip/journal/jcp/107/2/10.1063/1.474405,PhysRevA.47.979,PhysRevLett.76.1039,PhysRevA.57.4219}
\begin{equation}\label{cumulant_exp}
\gamma_{rspq}=\lambda_{rspq}+\gamma_{rp}\gamma_{sq}-\gamma_{sp}\gamma_{rq}
\end{equation}
 to evaluate a two-body density matrix which is not explicitly computed in DFT or MP2.
 Both one- and two-body density matrices are necessary  to calculate $\Sigma^{full}_1$ from Eqs.~\ref{sigma1}--\ref{G3}. The overall cost of evaluating $\Sigma_1$ with the factorized two-body density matrix is $O(n^5)$, where $n$ is the number of orbitals. This cost can be further reduced to $O(n^4)$ by employing density fitted Coulomb integrals. Multiple approximations to the expression for $\Sigma^{full}_1$  that can reduce the computational cost further are possible.  
 
 For a realistic systems  the exact $\Sigma^{full}_1$ can be approximated using the following scheme:
 \begin{center}
\bf{Iterative Scheme II}
\end{center}
\begin{enumerate}
\item calculate $\gamma^{(1)}$ from DFT or perturbation theory
\item calculate  $\gamma^{(2)}$ from cumulant expansion in Eq.~\ref{cumulant_exp}
\item calculate $\Sigma^{full}_1$ from Eqs.~\ref{sigma1}--\ref{G3}
\item calculate $U_{ijkl}$ for a subset of correlated orbitals 
\item calculate new $\gamma^{(1)}$ from the Green's function calculated using $U_{ijkl}$ 
\item calculate total energy
\item go to point 2 until convergence
\end{enumerate}
Obviously, such a procedure is not reliable for larger distances where the cumulant expansion breaks down since the two-body cumulant cannot be simply neglected. 

Here, again for calibration purposes, we avoid analyzing the embedding error and the errors  present are due to the use of effective integrals calculated by employing the cumulant expansion from Eq.~\ref{cumulant_exp}. To this end we use the one-body density matrix from FCI, while the two-body density matrix is constructed from Eq.~\ref{cumulant_exp}.

\begin{figure}[her!]
\centering
\includegraphics[width=1.0\columnwidth]{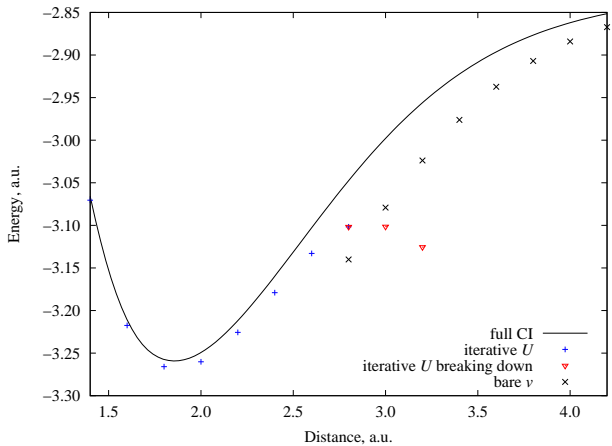}
\caption{\label{iter_apprx} Total energies for H$_{6}$ in the STO-6G basis set. On-site $U$ is obtained via an iterative procedure based on approximate $\Sigma^{full}_{1}$.}
\end{figure}

We have carried out a computational test for the same H$_{6}$ ring as in the previous sections with FCI using the STO-6G basis with on-site integrals only, thus making $U_{iiii}$ the only adjustable
parameter. To initialize the procedure bare $v_{iiii}$ integrals are used in the Hamiltonian to produce the starting FCI $\gamma^{(1)}$, and $\gamma^{(2)}$ was evaluated using Eq.~\ref{cumulant_exp}. The resulting energies are given in Fig.~\ref{iter_apprx}. 
From Fig.~\ref{iter_apprx} it is apparent that this iterative scheme can only work well as long as the approximation (\ref{cumulant_exp}) is valid. At $d = 3.0$ a.u. (\ref{cumulant_exp}) breaks down thus
leading to energies worse than those from the bare $v$ integrals. A clear sign of such breakdown is a situation when $U > v$, as opposed to the tendency observed in calculations
based on exact $\Sigma^{full}_{1}$ (\textit{cf.} Fig~\ref{screened}). 
Consequently, the iterative scheme based on employing the cumulant of the two body matrix is successful when only weak interactions are present. The frequency dependent
self-energy is then recovered equally well as in the examples above.   

\subsection{Approximating $\Sigma^{full}_1$ using the frequency dependent Green's function method}

\begin{figure}[her!]
\centering
\includegraphics[width=1.0\columnwidth]{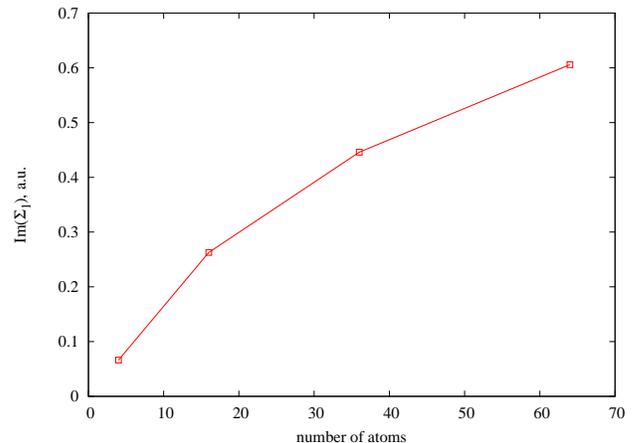}
\caption{\label{hydplaq} Values of $\Sigma^{full \ GF2}_1$ for the central atoms of hydrogen atom plaquettes with $n$ atoms.}
\end{figure}

Since the high frequency tail of $\Sigma(\omega)$ is describing short-time behavior, it is reasonable to expect  a perturbative method will recover the self-energy in the high frequency limit very well.  Indeed, our experience from analyzing 2D Hubbard models~\cite{Zgid12} confirms that second order iterated Green's function theory (GF2)~\cite{:/content/aip/journal/jcp/140/24/10.1063/1.4884951,Dahlenjcp2005} recovers the self-energy in the high frequency limit very well despite missing important features for low frequencies.
Since perturbative methods such as GF2, RPA or GW~\cite{HedinGWpra1965} can be performed with a very moderate cost for many molecular systems, the whole system can be treated to get $\Sigma^{full \ PT}_1$ which approximates the $\Sigma^{full \ FCI}_1$ very well. 
For solids, at least for insulators, one can perform GF2, RPA or GW on a large cluster embedded in a crystal lattice and expect the convergence of $\Sigma^{full \ PT}_1$ with the cluster size, thus avoiding performing the perturbative calculation on the whole system. 
We performed such a calculation using GF2 and employing a series of $N\times N$ hydrogen plaquettes, where $N$ is the number of atoms at the edge of the plaquette. These are designed to approximate a 2D solid hydrogen lattice.
From Fig.~\ref{hydplaq}, we can conclude that a converged $\Sigma^{full \ GF2}_1$ can be obtained for larger plaquette sizes.
From $\Sigma^{full \ GF2}_1$ the effective integrals can be evaluated using {\bf{Iterative Scheme I} }and replacing the quantities that come from FCI by those evaluated at the GF2 level. 

\section{Conclusions}\label{conclusions}

We performed multiple calibrations of a procedure for finding effective integrals based on the high frequency expansion of the self-energy. 
This scheme is different from other commonly used procedures for finding effective interactions, because it does not involve the construction of a  model Hamiltonian that is supposed to recover the most important low energy physics of the full problem. Instead, we construct a fictitious Hamiltonian that is parametrized such that the high frequency behavior of the full system can be recovered in an embedding calculation.
We discovered that the electronic energies are recovered very well by this procedure, resulting in a huge improvement of electronic energy when the fictitious system is parametrized with effective integrals rather than bare ones.
 While an ideal application of our prescription is to embedding methods such as DMFT, here we aimed to calibrate only the error coming from choosing the effective interactions, not the embedding error arising from choice of embedding method. 
 From our calculations it became apparent that the effective integrals are mathematical artifacts caused by incorporating the neglected non-local interactions in the fictitious local Hamiltonian used to evaluate the correlated Green's function. We have also observed that when multiple orbitals are embedded there is no unique parametrization for effective integrals but all of the parametrizations lead to good electronic energies. Consequently, in our method as long as the traditional effective integrals, such as the on-site $U$ and inter-site $J$ frequently used in DMFT or DFT+U calculations, are chosen to approximate the high frequency tail of the full system well, one can expect good results.

We have also analyzed the self-energy for the calculations with effective integrals. As expected, these self-energies were approximating the full system self-energy very well in the high frequency limit. For lower frequencies, the agreement between the FCI self-energy  and the one calculated with effective integrals was quantitative for small and completely stretched bond distances and qualitative for the intermediate case.

We have also discussed approximate ways of obtaining a high frequency expansion matrix $\Sigma^{full}_1$ based either on a cumulant expansion or Green's function perturbation theory. Perturbative Green's function methods may prove very useful and robust for evaluating effective interactions that can later be used by many methods for evaluating Green's functions which are working more efficiently with a specific type of the interaction structure. Since there is a freedom of how the $\Sigma^{full}_1$
can be parametrized, many effective interactions which are of the form $U_{ijij}n_{i\uparrow}n_{j\downarrow}$ can be successfully found. These are particularly suitable for the continuous time quantum Monte Carlo (CT-QMC) Green's function solver~\cite{RevModPhys.83.349} which is very successful in the condensed matter physics community.

\section{Acknowledgements}
D.Z. and A.R.  would like to acknowledge the DOE grant No. ER16391.
%

\end{document}